\documentclass[pre,showpacs,superscriptaddress,twocolumn]{revtex4}
\usepackage{epsfig}
\usepackage{dcolumn}
\usepackage{bm}

\begin{document}

\title{Diffuse interface approach to brittle fracture}
\author{V. I. Marconi and E. A. Jagla}
\affiliation{The Abdus Salam ICTP, Strada Costiera 11, (34014) Trieste, Italy}
\date{\today}

\begin{abstract}

We present a continuum model for the propagation of cracks and fractures in brittle materials.
The components of the strain tensor $\varepsilon$ are the fundamental variables.
The evolution equations are based on 
a free energy that reduces to that of linear elasticity for small $\varepsilon$,
and accounts for cracks through energy saturation at large values of $\varepsilon$.
We regularize the model by including terms dependent on gradients of $\varepsilon$ in the free energy.
No additional fields
are introduced, and then the whole dynamics
is perfectly defined. 
We show that the model is able to reproduce basic facts in fracture physics, like the Griffith's
dependence of the critical stress as a minus one half power of the crack length.
In addition, regularization makes the results insensitive to the numerical mesh used,
something not at all trivial in crack modeling.
We present and example of the application of the model to predict the growth and
curving of cracks in a non-trivial
geometrical configuration.
\end{abstract}
\maketitle

\section{Introduction}

There are many problems in condensed matter physics and materials science in which the aim is to
describe sharp interfaces separating regions with qualitatively different properties. 
This occurs for instance in solidification, dendritic growth, solid-solid transformations, grain
growth, etc. 
Traditionally,
two approaches have been followed to tackle these problems. 
In one of them, the problem is solved for a fixed position
of the interface, and based on this, the expected evolution of the interface in the following 
time step is calculated, and the process repeated. This method has the practical drawback
that the different structure of the interface at each time step makes necessary the
full solution of a new problem each time. The second approach is sort of brute force, in which 
the system is modeled to the atomic scale, and evolved according to their
(Newtonian) equations of motion. The problem with this approach is that it
is impossible in practice to span the many orders
of magnitude between the atomic scale and the relevant macroscopic 
scale.

The diffuse interface technique (including the so-called phase field models)
is a novel powerful approach to study this kind of problems\cite{reviews}.
It typically describes the sharp interface by an additional 
field $\phi$ (or more than one). For
instance in a solidification problem $\phi$ can be taken to be 
0 in the liquid and 1 in the solid. If the spatial 
variation of $\phi$ is known the interface can be located. Then the problem of keep tracking
of the interface is eliminated against having to include also $\phi$ 
as a new dynamical variable. $\phi$ is coupled (usually phenomenologically) 
to the original degrees of freedom of the problem, and its
dynamic evolution is not defined {\em a priori}, but has to be seeded from outside.

A key ingredient in phase field models is {\em regularization} of the field $\phi$. Although the
sharp interface is the sensible case, in order to implement the theory a smooth transition of $\phi$ 
between the values on both sides of the interface is necessary. Then the interface acquires a fictitious
width, which however does not alter the physical behavior of the system if
it is much smaller than any other relevant length scale.
An additional, very important effect of regularization is to make the properties of the system
independent of the underlying numerical mesh used to implement the problem on the computer.
Regularization is usually achieved by the inclusion in the theory of terms that depend on gradients
of $\phi$, penalizing rapid spatial variations of this quantity.

Within the field of fracture, the phase field models that have been proposed 
include those of Aranson, Kalatsky and Vinokur \cite{aronson}, 
Karma, Kessler and Levine \cite{karma}, and Eastgate {\it et al.} \cite{sethna} (see also  \cite{kacha}). 
All of them use an additional scalar field $\phi$ as a phase field, that is taken to be (asymptotically) 
zero within fractures and one within the intact material.

There is now a general consensus that a complete description of the fracture process cannot
be given only in terms of macroscopic variables. In fact, the divergence of the stress field
near to the crack tip implies that physical conditions change a large amount on
distances of the order of interatomic separation. Then details of the material at the atomic scale
can have an effect on the macroscopic behavior of cracks. On the other hand, the roughly similar 
phenomenology of
crack propagation observed in very different materials raises the expectation that a general description with
a minimum amount of parameters dependent on microscopic characteristics is feasible. This is in the
spirit of the phase field approach to fracture: one may think that the microscopic variables have an effect
that translates in the form given to the energy density of the phase field, in the form of the
terms coupling the phase field to the elastic degrees of freedom, and in the dynamics assumed for it.
Except in this effective way, microscopic parameters do not appear in the phase field 
formalism.

The phase field approach is already giving promising results. For instance, it has been shown that
crack instabilities, oscillations and bifurcation can be obtained within this scheme \cite{karma04}.
The sharp interface limit of some phase field models of stress induced instabilities has been studied in 
\cite{kessner}. Its possible relevance to fracture is
given in  \cite{sharpin}.

We are going to present here a diffuse interface approach that has some qualitative 
difference with previous ones. Most importantly, it does not 
introduce additional variables into the problem, the full set of variables are 
the components of the strain tensor $\varepsilon$\cite{nota0}.
Description of fracture is achieved by the nonlinear form of the effective free energy density
as a function of 
$\varepsilon$. Actually, our energy is quadratic for small displacements (and then
correctly describes
linear elasticity) and
saturates for very large displacements, then describing fracture
(the saturation energy value being related to the fracture energy). Regularization is provided by terms
in the free energy of the generic form $(\partial \varepsilon)^2$. 

There are a number of reasons to pay attention to this model, 
both conceptual and from the point of view of implementation.
First of all, the absence of additional degrees of freedom makes this model be probably the
simplest continuous (non-atomistic) description of the fracture process. 
It is then interesting to know how, and to what extent, fracture phenomenology is captured 
by the model.

From a practical perspective there are two important things to point out.
First, an important characteristic is the tensorial nature of the variable describing the occurrence
of fractures. In the approaches in which a scalar field $\phi$ is introduced, knowing that 
$\phi$ has become
zero at some point tells that a fracture is passing through that point, but does not tell anything about what
direction the fracture follows. In our case fracture is described by $\varepsilon$ itself, and then if we know that a
fracture is passing through some point, we can say immediately which direction it has. We believe this is an
important computational advantage: a single cell of a computational mesh is sufficient to encode the information
about existence and direction of fracture. In models using a scalar field $\phi$ we need a whole neighborhood of
a computational cell to encode the same information. Second, previous attempts to model
fracture through non-linear elasticity have used typically 
the displacement field as fundamental variable. For those
theories regularization is a problematic 
issue as it typically leads to higher order differential equations which are difficult to solve numerically. 
Our equations contain only second order derivatives of the strain field, and
thus they much more smooth to solve numerically.

In this paper we concentrate mainly in the presentation and validation of the model, giving only a short
presentation of a non trivial application. 
We have divided the presentation in the following form. In the next Section we define the model and give
analytically the structure on an infinite straight crack. Section III shows in detail that the model 
is able to reproduce the Griffith's criterion. In Section IV we present an example in which the crack paths
are determined in a non trivial geometric configuration. In Section V we give a perspective of our planned 
future work with the model.

\section{The basic model}

The fundamental variables of the model are the components 
of the strain tensor $\varepsilon$:

\begin{equation}
\varepsilon_{ij}\equiv \frac{1}{2}\left(\frac{\partial u_i}{\partial x_j}+
\frac{\partial u_j}{\partial x_i}\right)
\label{uno}
\end{equation}
where $u_i({\bf r})$ are the displacements with respect to the unperturbed positions  
\cite{nota1}.
Our approach follows closely that  in Ref. \cite{shenoy} (see also \cite{kartha}) where it was used 
to study textures in ferroelastic materials and martensites. For
clarity we present the model for a two dimensional geometry, 
the generalization to three dimensions being conceptually straightforward,
although of course more involved.
The symmetric tensor $\varepsilon_{ij}$ has three independent components. 
For convenience we will choose them in the form
\begin{eqnarray}
e_1&\equiv&(\varepsilon_{11}+\varepsilon_{22})/2\nonumber\\
e_2&\equiv&(\varepsilon_{11}-\varepsilon_{22})/2\label{dos}\\
e_3&\equiv&\varepsilon_{12}=\varepsilon_{21}\nonumber
\end{eqnarray}
which are named respectively the dilation, deviatoric and shear components.
These three variables 
are however not independent. In fact, since $\varepsilon$
is derived from the two displacements $u_1$, $u_2$, there is one constraint that has to be fulfilled, leaving
only two independent variables. The constraint is known as the St. Venant condition \cite{stv}, 
and it can be written as

\begin{equation}
(\partial^2_x+\partial^2_y)e_1-(\partial^2_x-\partial^2_y)e_2-2\partial_x\partial_y e_3=0.
\label{stv}
\end{equation}
It is easy to show that this is an identity if the definitions (\ref{uno}) and (\ref{dos}) are used\cite{nota}.

To define the model we need to know the form of the local free energy density $F_L(\varepsilon)$. 
To correctly describe elasticity for small displacements the limiting form $F_L^0$ of $F_L$ for 
small $\varepsilon$ should be

\begin{equation}
F_L^0(\varepsilon)= \frac12C_{ijkl}\varepsilon_{ij}\varepsilon_{kl}
\end{equation}
where $C_{ijkl}$ is the fourth rank tensor of elastic constants of the material.
We will specialize the expressions to an isotropic material, as this was our first aim 
to use this kind of model.
In the isotropic case $F_L^0$ reads

\begin{equation}
F_L^0(\varepsilon)= 2B e_1^2 +2\mu \left( e_2^2+e_3^2\right )
\label{iso}
\end{equation}
where $B$ and $\mu$ are the two dimensional bulk and shear modulus of the material (related to the 
three dimensional values $B^{3D}$ and $\mu^{3D}$ by
$\mu=\mu^{3D}$, $B=B^{3D}+\mu^{3D}/3$ for the case of plane strain, and $\mu=\mu^{3D}$,
$B=3B^{3D}\mu/[B^{3D}+4\mu/3]$ for the case
of plain stress).

The previous expression of the free energy must be extended to large strains to account for fracture. 
To do this the energy must saturate for large deformation. This is the main requirement, and different 
choices can be made \cite{nature}.  In the simulations presented below for isotropic materials 
we have chosen the form

\begin{equation}
F_L(\varepsilon)=\frac{F_L^0(\varepsilon)}{1+F_L^0(\varepsilon)/f_0}
\end{equation}
The limiting value $f_0$ of $F_L$ for $\varepsilon \rightarrow\infty$ 
is the energy density necessary to impose locally a very large value for at least one component
of $\varepsilon$, and it is then
related to the crack energy of the problem (see below). From the present free energy form, we
can say that a crack is nucleating when $F_L^0\sim f_0$, i.e., when
typical values of $e$'s are $e_i\sim (f_0/B)^{1/2}$ (assuming $\mu\sim B$, as it is usually the case).
Cracks in the system are thus characterized as regions where $F_L^0\gg f_0$.

The second crucial ingredient of the model is regularization. That is provided by gradient terms in the free
energy density. The terms we use are typically of the form
\begin{equation}
F_{\nabla}\equiv\sum_{i=1,2,3} \alpha_i (\nabla e_i)^2 
\label{alfbet}
\end{equation}
with numerical coefficients $\alpha_i$. To retain rotational invariance we have to choose 
$\alpha_2=\alpha_3$. In certain cases, and to avoid some unphysical behavior, these terms have to be
 cut off at large values of $e_i$ (see the next section for justification and details), and thus
the gradient part of the free energy density is chosen as
\begin{equation}
F_{\nabla}=\sum_{i=1,2,3} \alpha_i (\nabla e_i)^2 s(e_i)
\label{fnabla}
\end{equation}
where $s(e_i)$ goes to 1 for small $e_i$, and tends to zero at large $e_i$. The
actual form of $s(e_i)$ we use is

\begin{equation}
s(e_i)=\frac{1}{1+(F_L^0/f_1)^\kappa}
\label{cf}
\end{equation}
where $F_L^0$ is given in (\ref{iso}), $f_1$ is a cut off value, 
and the exponent $\kappa$ controls the sharpness of the cut off (see next section).

Once the full free energy is defined, the equations of motion are obtained  by including a
kinetic term $T$, which is quadratic in temporal derivatives of the displacements (i.e.,
$T\sim \frac 1 2 \rho \dot{\bf u}^2$, for some generic density $\rho$) and then it has to be transformed 
to a function of $e_i$
(see \cite{shenoy} for the details).  The equations to be solved
are more conveniently written in Fourier space (we use $E_i$ for the Fourier transforms of $e_i$, and
$F=F_L+F_{\nabla}$).
The result is
\begin{eqnarray}
\frac{4\rho}{k_x^2k_y^2}(k^2 \ddot E_1+(k_x^2-k_y^2)\ddot E_2)&=&%
-\frac{\delta F}{\delta E_1^*}-A_1\dot E_1 +k^2 \Lambda\nonumber\\
\frac{4\rho}{k_x^2k_y^2}((k_x^2-k_y^2)\ddot E_1+k^2\ddot E_2)&=&%
-\frac{\delta F}{\delta E_2^*}-A_2\dot E_2 +(k_y^2-k_x^2) \Lambda\nonumber\\
0&=&-\frac{\delta F}{\delta E_3^*}-A_3\dot E_3 -2k_x k_y \Lambda \label{ocho}
\end{eqnarray}
where $A_1$, $A_2$, $A_3$  are phenomenological damping coefficients (that we will 
take to be equal $A_i\equiv A$) and
$\Lambda$ is a Lagrange multiplier depending implicitly on $E$'s, chosen to 
enforce at every moment the St. Venant constraint, that
in Fourier space reads
\begin{equation}
(k_x^2+k_y^2)E_1-(k_x^2-k_y^2)E_2-2k_x k_y E_3=0.
\end{equation}

In the examples below, the dynamic equations (\ref{ocho}) are solved on a rectangular numerical 
mesh, using periodic boundary conditions in systems up to sizes of 512 $\times$ 512. In some
cases we work with rectangular samples, with the length perpendicular to the crack being two or three
times that along the crack, as we have observed that finite size effects are lower in this configuration
than in a square sample with the same area.
Also, the dynamic equations are solved in the overdamped regime, in which the second 
order time derivative terms in (\ref{ocho}) are neglected.

\subsection{Infinitely long, straight crack. Analytical results}

As a necessary starting point, and also because it is probably the only case that can be
treated analytically,
we present here the structure of an infinite, straight crack in our model.
In this geometry, all quantities depend only on the coordinate
$x$ (the crack is assumed to lie along the $y$ direction), and the model becomes effectively one-dimensional.
In fact, assuming the boundary conditions impose no strain in the $y$ direction, we have
$e_1(x)=e_2(x)$, $e_3=0$. As expected, there is a single order parameter in this configuration, that can be
taken to be $e_1$. The free energy of the model restricted to the present case
becomes simply 

\begin{equation}
F/L_y=\int  \frac{B^*e_1^2}{1+B^* e_1^2/f_0}dx+{\alpha} \int |\nabla e_1|^2 s(e_1)dx
\label{bici}
\end{equation}
where $L_y$ is the system size along the $y$ direction, $B^*=2B+2\mu$,
$\alpha=\alpha_1+\alpha_2$, and the cut off function $s$ is now a function of $e_1$ alone.
For the choice in Eq. (\ref{cf}) we get now 
\begin{equation}
s(e_1)=\frac 1{1+(B^*e_1^2/f_1)^{\kappa}},
\label{s1d}
\end{equation}
It is clear that for $\alpha=0$ the solution to Eq. (\ref{bici}) has $e_1=0$ except at a single point
$x_0$. The position $x_0$ is undetermined. This solution describes
the system broken at $x_0$. The fracture energy per unit length $\gamma$ in this case is non-zero 
only if a finite discretization $\delta$ is used. In this case $\gamma= f_0 \delta$. 

\begin{figure}
\centerline{\epsfxsize=8.cm \epsfbox{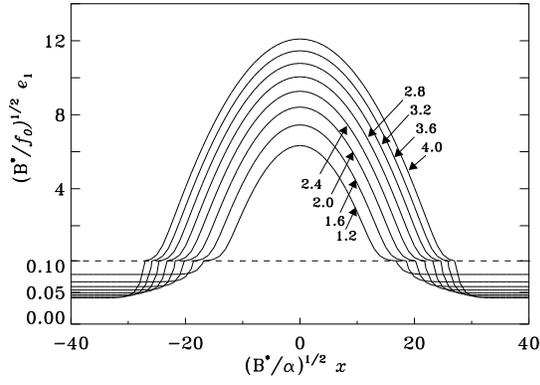}}
\caption{Profiles of $e_1$ for the one-dimensional problem, at different values of 
the global strain $\bar e_1=\delta L_x/L_x$. 
Labels are the values of $(B^*/f_0)^{1/2}\bar e_1$ used for each curve.
Note that $e_1$ does not go to zero away from the fracture,
but a remmanent strain remains. This value however, decreases when the 
fracture is opened wider.}
\label{f1}
\end{figure}

To work out the finite regularization case ($\alpha\ne 0$),
it is more convenient to solve the problem under the assumption of an applied stress 
$\sigma$ along the $x$ direction. To do this we have to minimize the stress dependent free energy $F_{\sigma}$,
given by
\begin{equation}
F_{\sigma}/L_y=F/L_y-2\sigma\int  e_1(x)dx
\label{fsigma}
\end{equation}
(the factor of two is due to the fact that in the present case 
$e_1=\frac 12\frac{\partial u_x}{\partial x}$).
Let us consider first the case $f_1\rightarrow \infty$, which implies (according to (\ref{s1d}))
$s(e_1)\equiv 1$, i.e., no cut off of the gradient terms at 
high values of $e_1$.
As in any one dimensional mechanical problem, the solution of Eqs. (\ref{bici}),
(\ref{fsigma}) 
is reduced to the evaluation of an integral.
The numerical integration gives the profiles shown in Fig. \ref{f1}.
An important observation is that now the whole profile is smooth, and this implies that all macroscopic
parameters that are numerically calculated will be independent of the mesh discretization $\delta$
if this is small enough. 

For low $\sigma$, in the region where $e_1\gg \equiv\sqrt{f_0/B^*}$ the following analytical solution is obtained

\begin{equation}
e_1(x)=e_1^{max}-\frac{\sigma}{2\alpha}x^2
\label{e1}
\end{equation}
where $e_1^{max}$ is order $f_0/\sigma$.
The sharp fracture of the case $\alpha=0$ transforms now into a smooth object that occupies a finite width $w$
in the system. 
From (\ref{e1}) this width can be estimated to be 
\begin{equation}
w\sim \sqrt{\alpha f_0}\sigma ^{-1}.
\label{w}
\end{equation}
Note that $w$ measures the width of the fracture in the {\em original},
unstrained reference system.
An important parameter to be calculated is the opening of the fracture $\Delta$, 
defined as $\Delta\equiv\int \left(e_1(x)-e_1(\infty)\right) dx$. This has the meaning of the strain that the
system is able to accommodate due to the existence of the crack. Its main contribution
for low $\sigma$ comes from the central part of the crack, described by Eq. (\ref{e1}). We get
\begin{equation}
\Delta\sim f_0^{3/2}\alpha^{1/2}\sigma^{-2}
\end{equation}
and from here and (\ref{w})
\begin{equation}
w\sim \alpha^{1/4}f_0^{-1/4}\Delta^{1/2}
\end{equation}
Then we see that upon increasing the opening $\Delta$ of the fracture, the width $w$ increases 
as $\Delta^{1/2}$, and the stress decays only as $\Delta^{-1/2}$. Then this model crack 
relieves the stress in the system only asymptotically. 
However, the decaying of the stress with $\Delta$ is too slow to give a finite crack
energy.
In fact, the crack energy per unit length $\gamma$ (defined as the energy of the system with the crack, minus
the energy of the system at the same stress without the crack) can be estimated as:
\begin{equation}
\gamma \sim f_0 w\sim \alpha^{1/4}f_0^{3/4}\Delta^{1/2}
\label{gamma}
\end{equation}
This means that the present regularization scheme does not
provide a crack with a finite energy in the limit of large stretching, i.e., 
$\Delta\rightarrow \infty$. 

In the next section we will argue that this behavior does not invalidate completely the use of the present
algorithm with $f_1\rightarrow\infty$ if we are interested in cracks that do not become infinitely long.
Notwithstanding, if we want a more accurate description of cracks, an implementation in which the 
energy density of the crack remains finite as its length diverges is mandatory. The
following is a possibility to obtain this.

In order to have a model that relieves more efficiently the elastic energy, 
we first remind that
gradient terms are necessary for a fracture to
propagate without interference of the numerical mesh, therefore, they are important 
only when fracture is forming, namely,
when typical values of $e_i$ are close or lower than $\sim(f_0/B)^{1/2}$.
In the regions in which the fracture has nucleated, typical values of 
$e$'s become much larger, and the gradient terms are 
not necessary any more. Then we can
weaken their effect in those regions by choosing a non-trivial cut
off function as given by Eq. (\ref{cf}) (or Eq. (\ref{s1d}) in the present one-dimensional case). It should be
emphasized that this {\em ad hoc} modification of the free energy 
does not introduces new parameters in the relevant regions that determine 
crack growth, namely, close to the crack tips.

\begin{figure}
\centerline{\epsfxsize=8.cm \epsfbox{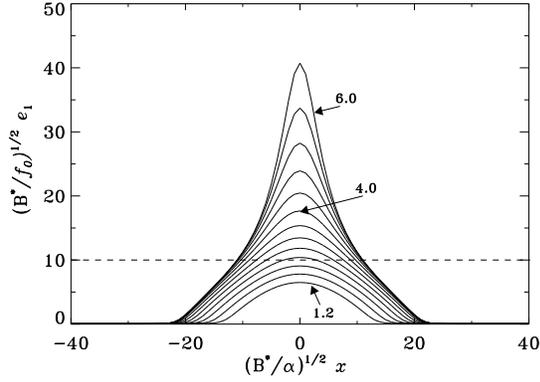}}
\caption{Same as previous figure when the gradient terms are cut off
at large values of $e_1$, as explained in the text. Parameters used are
$f_1/f_0=100$, $\kappa=2$. 
Note that contrary
to the case in the previous figure, the fracture
width (and this implies also the fracture energy) saturates to a finite value
for large $\bar e_1$.}
\label{f2}
\end{figure}

Our aim is to choose a value of $\kappa$
that generates cracks with finite width and energy in the limit $\Delta\rightarrow \infty$.
In order to do this, we first notice that for $e_1\gg \sqrt{f_0/B^*}$ (i.e., where elastic forces
become vanishingly small), the Euler-Lagrange equation corresponding to 
Eqs. (\ref{bici}), (\ref{s1d}), and (\ref{fsigma}) leads to
\begin{equation}
\frac{\alpha}{1+\left(\frac{B^*e_1^2}{f_1}\right)^{\kappa}}\left(\frac{de_1}{dx}\right)^2+2\sigma e_1=\mbox{constant}  
\label{22}
\end{equation}
In particular, if $f_1\rightarrow \infty$, we re-obtain from here the behavior given in (\ref{e1}).
The constant in Eq. (\ref{22}) must be calculated matching the present solution for $e_1\gg \sqrt{f_0/B^*}$,
with that for $e_1\lesssim \sqrt{f_0/B^*}$. Assuming $f_1\gg f_0$ for simplicity, it is obtained that this constant is
$D f_0$, where $D$ is a (order 1) numerical factor. Upon integration, Equation (\ref{22}) 
allows to find the full 
profile of the crack
(in Fig. \ref{f2} we can see the results of numerical integration for $\kappa=2$).
We concentrate in the case $\sigma\rightarrow 0$, in which the crack has relaxed completely the applied
stress.
The crucial result is that in this case, and for $\kappa>1$, it exists a 
well defined profile of the crack, given by
\begin{equation}
e_1^{\sigma=0}(x)=\sqrt\frac{f_1}{B^*}g^{-1}\left(\sqrt{\frac{f_0B^*D}{f_1\alpha}}|x|\right)
\end{equation}
where the function $g$ is defined as
\begin{equation}
g(u)=\int_u^{\infty} \frac{dw}{\sqrt{1+w^{2\kappa}}}.
\end{equation}
For $x\rightarrow 0$ the limiting form is
\begin{equation}
e_1^{\sigma=0}(x\rightarrow 0)=\sqrt{\frac{f_0}{B^*}}\left(
(\kappa-1)\sqrt{\frac{f_0B^*D}{f_1\alpha}}|x|\right)^\frac{1}{1-\kappa}
\label{23}
\end{equation}

The opening of the crack $\Delta=\int  e_1(x)dx$ gives a finite value if $\kappa>2$.
This means that the system has completely relaxed the applied stress with a finite opening of
the crack. Any further increase $\Delta ^*$ of $\Delta$ is accommodated in the system at the center of the crack,
through a singular term $\Delta^* \delta(x)$ added to (\ref{23}).
For $1<\kappa<2$, the value of $\Delta$ is divergent due to the non-integrable divergence of $e_1$ around the 
origin in (\ref{23}). In this case the stress is relaxed only asymptotically, 
but still rapidly enough (compared
to the $f_1\rightarrow \infty$ case) to guarantee a finite width $w$ and energy $\gamma$ of the fracture.
In fact, from (\ref{23}) 
the width $w$ of that part of
the system for which $e_1\gtrsim \sqrt{f_0/B^*}$ can be estimated to be 
\begin{equation}
w\sim\sqrt\frac{f_1\alpha}{f_0 B^*}
\end{equation}
in the same way, the energy of the crack $\gamma$ becomes finite, and its value is proportional to
\begin{equation}
\gamma\sim f_0 w\sim \sqrt\frac{f_0f_1\alpha}{B^*}
\end{equation}

From a numerical point of view, the algorithm with $\kappa>2$ is probably too singular
to be implemented successfully.
We have implemented the case $\kappa=1.5$, which we have found is 
numerically tractable, and provides and almost 
perfect verification of the Griffith's criterion 
even in the quite small system sizes that we are
using.

\section{Finite straight crack under model I loading: Verification of the Griffith's law}

One of the basic cornerstones of fracture physics, described in the very first pages of any fracture book,
is the so called Griffith's criterion (GC). It states that under the application of a remote stress $\sigma$ on
a (effectively two-dimensional) system with a preexistent crack of length $l$, 
the crack will extend and eventually break the sample if
$\sigma$ is larger than some critical value $\sigma_{cr}$ which scales as $l^{-1/2}$.
A simple justification of this behavior can be given on energetic grounds. Upon the application of the stress
$\sigma$, the system with the crack stores an elastic energy that is reduced in a quantity $E_{el}=C \sigma^2 l^2$
with respect to that of the system without the crack
($C$ is proportional to an elastic constant of the material). On the other hand, the creation of a crack of length $l$
is assumed to have an energy cost of $E_{cr}=\gamma l$, where $\gamma$ defines the fracture energy per unit length.
The total energy of the system as a function of $l$ is then given by
\begin{equation}
E=-E_{el}+E_{cr}=-C\sigma^2 l^2+\gamma l
\label{e}
\end{equation}
and it has a maximum as a function of $l$ at $l_{max}=\gamma/(2C\sigma^2)$. If $l>l_{max}$, the system relieves sufficient
elastic energy upon crack length increase to pay for the energy cost of crack creation, and the crack
typically extends abruptly and breaks the material. If $l<l_{max}$ there is no sufficient energy in the system
to make the crack enlarge. Note that according to (\ref{e}), the system would prefer to reduce $l$ in order to
reduce its energy. Experimentally this `healing' of the crack is prevented by irreversible processes: the crack
energy is not recovered upon crack healing, and then the situation is that any $l<l_{max}$ is a stable crack.
The previous arguments do not depend on the orientation of the crack in the system, assuming the 
material and the remote applied stress are isotropic.
The GC is at the base of the fracture of brittle materials, and has to be satisfied by
any model devised to describe such process. 

In our simulations we control the mean uniform strain $\bar e_1$. This correspond to an isotropic stress
$\sigma=2B\bar e_1$.
In order to correctly calculate the critical stress $\sigma_c$ 
for cracks of a given length, we start from an initial spatial
distribution of the variables $e_1$, $e_2$, $e_3$ that roughly describes a crack, 
and apply a stabilization
procedure through a negative feedback loop in the program, that monitors the length of the crack (defined 
using the contour defined by the value of the 
elastic energy $F_L^0$=2B), and reduces (increases) the applied strain when the length increases
(decreases). Results in Figs. \ref{f3}, \ref{f4}, and
\ref{f7} below, where obtained with this procedure.

\begin{figure}
\centerline{\epsfxsize=8.cm \epsfbox{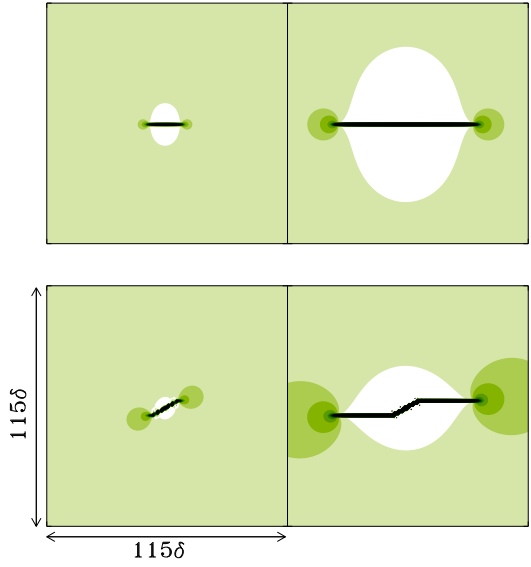}}
\caption{(Color online) Effect of the numerical mesh on the propagation of cracks in the model without
regularization ($\alpha=0$). Snapshots of cracks of original length $l_0=16 \delta$,
growing with $\theta=0 \deg$ (upper plots), and 
$\theta=30 \deg$ (lower plots) 
and a remote isotropic strain $\bar e_1=0.19$, larger than the corresponding critical value.
The left panels are the configuration soon after the application of the
loading, and the right panels
are those some time later.
The influence of the numerical mesh is
evident. 
The full simulated system size is $256\times256$.
The key to the gray (green) scale here and
in all following figures is as follows:
We plot the values of $e_1$ from brighter (lowest $e_1$) to darker color for a value
$e_1  {\protect \gtrsim} 2$ 
(corresponding to the crossover to cracked material). All values above that one 
are plotted black and correspond to the region inside the crack.
}
\label{f5}
\end{figure}

The model with no regularization 
satisfies the GC if we restrict to cracks running in a single direction 
with respect to the underlying numerical mesh. This is shown in Fig. \ref{f3}.
We also see  that no noticeable
system size effects are observed for the system sizes used. As we use periodic
boundary conditions this means that the crack is not influenced by the elastic field of its neighbor images.
However, the value of $\sigma_{c}$ is strongly dependent on the orientation of the crack as
 shown in Fig \ref{f4}.
This is one typical
drawback of many discrete models of fracture when trying to simulate isotropic materials. 
Moreover, the influence of the numerical mesh has a clearly visible
manifestation: as Fig. \ref{f5} shows, if a crack is placed at a finite angle with respect to the mesh, when the critical stress is
reached,
the crack opens along one of the lattice main directions, instead of extending along its original 
direction as it should do in an isotropic system.

The gradient terms are included in the model to solve this problem, and to make the behavior isotropic.
The question of the satisfiability of the GC in the presence of a non-zero $\alpha_i$ 
is not
trivial since as we showed in the previous section, the energy of the crack per unit length does not saturate 
upon increasing strain unless an adequate  cut off of the gradient terms is included.
Let us first study the effect of a finite  
regularization without cut off
($f_1\rightarrow\infty$ in Eq. (\ref{cf})). 
The fact that in this case the crack energy per unit length of an infinite crack is divergent (as seen in the previous
section) is a manifestation of the fact that the crack energy of a crack of {\em finite} length $l$ grows more rapidly
than $l$ itself. On the basis of the energetic arguments for the GC, we expect here a dependence
of $\sigma_{c}$ on $l$ of the form $\sigma_{c}\sim l^{-\beta}$ with $\beta<1/2$.

\begin{figure}
\centerline{\epsfxsize=8.cm \epsfbox{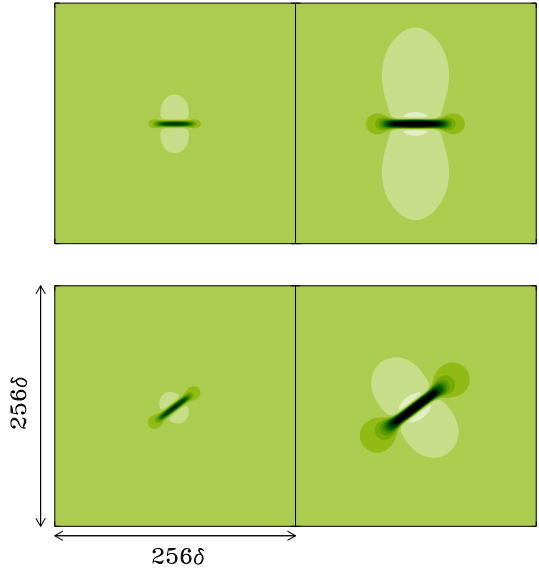}}
\caption{(Color online) Same as previous figure for $\alpha=0.5 B\delta^2$,  $f_1 \rightarrow \infty $, $l_0=40\delta$ and
 $\bar e_1=0.56$ .
 The effect of the numerical mesh is
undetectable.}
\label{f6}
\end{figure}

Numerical simulations first of all confirm that the behavior of the system can be made isotropic by including
regularization. In fact Fig. \ref{f4} shows that the critical stress become independent of the angle between the
crack and the numerical mesh. More than that, the crack extends along the original direction it had,
independently of the numerical mesh (see Fig. \ref{f6}).
However, as anticipated, a power low decaying with an exponent lower than 1/2 is obtained for the
critical stress as a function of crack length (see
Fig. \ref{f7}).
The fitted power for the parameters used is $\beta = 0.402 \mp 0.009$. 
Then the model with finite $\alpha_i$ but infinite $f_1$ gives a slightly
incorrect behavior of critical stress as a function of crack length. If this discrepancy can be considered small,
then the model with infinite $f_1$ (which is easier to implement) can be used. 
If the previous discrepancy is considered serious (whether it is serious or not will depend on the
particular problem under study) then a finite $f_1$ formalism has to be implemented. As already discussed, a
power $\kappa>1$ in Eq. (\ref{cf}) has to be used to guarantee that the GC is satisfied. 
In Fig. \ref{f7} we show results indicating that a very
good fitting to the GC is obtained for $\kappa=1.5$ and $f_1=14.6 B$. This value of $f_1$ was chosen to obtain the
best verification of GC in the finite system we are using, but in principle any (finite) value of
$f_1$ should reproduce the 1/2 power dependence in the case of sufficiently large system sizes.
Fig. \ref{f8} shows the three dimensional profiles of the crack with $l=83\delta$, stabilized  at the critical
stress.

\section{Elastic interactions between cracks}

\begin{figure}
\centerline{\epsfxsize=8.cm \epsfbox{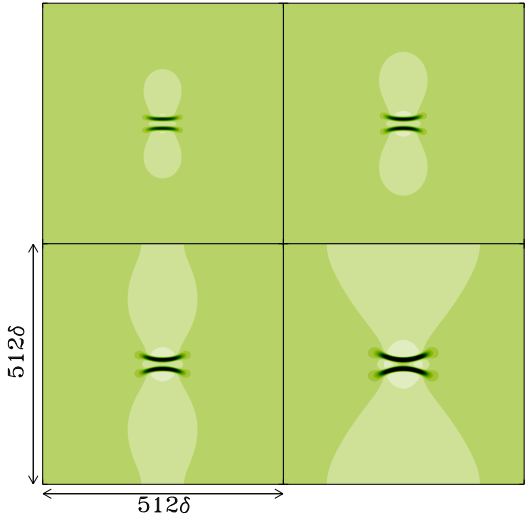}}
\caption{(Color online) Time sequence for two parallel cracks (length $l_0=60\delta$, separation equal to $20\delta$)
loaded isotropically under $\bar e_1$= 0.516 (larger than the corresponding critical value), 
with $\alpha= B\delta^2$, $f_1=14.6 B$, $\kappa=1.5$.
 The cracks extend with some divergence, due to a 
non-trivial effect of elastic interaction between them. 
}
\label{f9}
\end{figure}

\begin{figure}
\centerline{\epsfxsize=8.cm \epsfbox{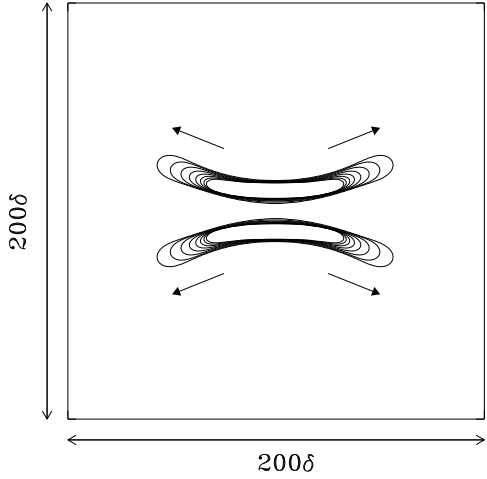}}
\caption{Time sequence in contours of the elastic energy at $F_L^0=2B$ for the parallel cracks of
the previous figure (equally spaced each $\Delta t=25A/B$, arrows indicate increasing time)}
\label{f10}
\end{figure}

The prediction of crack trajectories under general loading conditions for bodies of arbitrary shape and possibly
with pre-existent cracks is very important in 
many engineering applications.
The present formalism is well suited to study this kind of problems, in particular
in those cases in which slight deviations from straight
propagation are expected. These cases are particularly difficult to tackle with a non-regularized model.
As a very simple an illustrative example of that, we consider a pair of parallel cracks loaded
isotropically. We show here some qualitative results,
leaving a more detailed quantitative analysis for a forthcoming work.
We first stabilized the two parallel cracks by the feedback mechanism already explained, then
stop this stabilization, and increase a small amount the stress, and follow the crack
evolutions as they grow.
Snapshots of the system during crack propagation (Figs. \ref{f9}, \ref{f10}) show
that the cracks propagate diverging from straight 
propagation. This is a non-trivial
effect caused by the elastic interaction between the two cracks.
Note that in the present case, cracks propagate from {\em both} fractures, because of the
perfect mirror symmetry of the problem with respect to the middle plane.

\begin{figure}
\centerline{\epsfxsize=8.cm \epsfbox{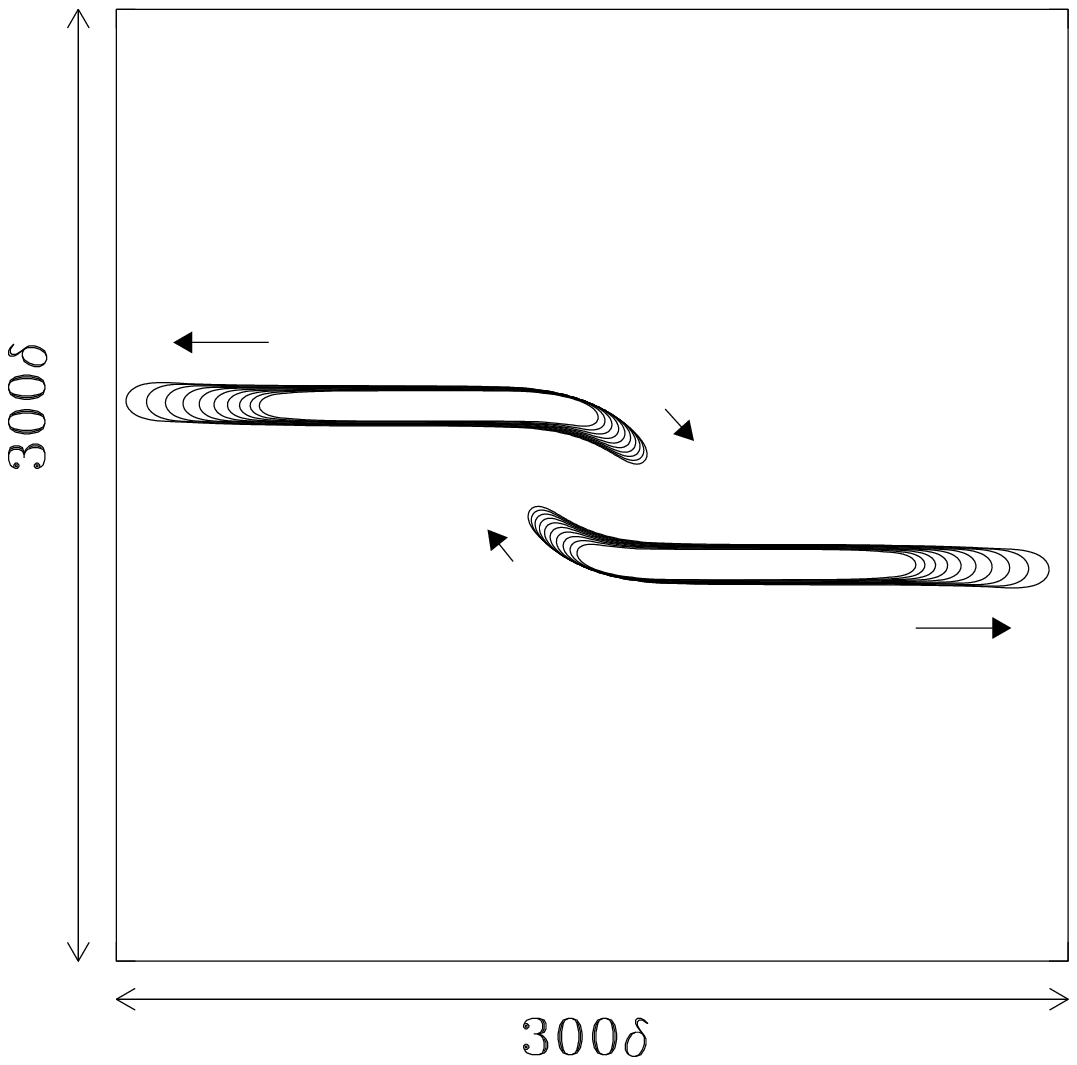}}
\caption{Same as previous figure for skew-parallel cracks with
initial length $l_0=100\delta$, perpendicular distance between them equal to $50\delta$ and horizontal distance
between the outer tips equal to $200\delta$ (simulated system size: $384\times 512$).  
The isotropic load is     $\bar e_1$=0.3, corresponding to a stress $\sigma =0.6 B$, larger 
than the corresponding critical value
$\sigma_c \simeq 0.5B$.}
\label{f11}
\end{figure}

In a slightly different configuration, we place the two parallel cracks shifted (Fig. \ref{f11}).
Now the propagation of cracks from the internal tips is strongly influenced by the nearby crack, producing 
a geometrical pattern of curved cracks that is well known (see for instance \cite{prob}). The propagation from the external tips is now almost not influenced by
the second crack and then essentially straight.

\section{Summary, outlook and conclusions.}

In summary, we have presented a model for the study of cracks propagating in brittle materials. 
The model does not use additional variables others than the strain tensor, and then 
is a minimalist continuum model for description of cracks.
As a crucial ingredient it includes 
regularization terms that make the cracks be smoothed. We have shown that in this 
form the model can describe accurately an isotropic material in
conditions in which the non-regularized model fails neatly. 

We have validated the model showing how it can fit accurately the Griffith's law for the critical stress
as a function of crack length. We have seen that in order to obtain this result the regularization
has to be softened in the interior of the cracks. This {\em ad hoc} modification however
leaves intact the model in the neighborhoods of the crack tips, where the processes responsible
for crack advance take place.
As a first application we have shown how the model can predict the propagation and eventual
bending of cracks induced by elastic interactions between them.

We believe that the present technique is straightforward to implement and computationally efficient, 
and that it addresses in a phenomenological way the very important applied problem
of predicting crack evolution,  without requiring as explicit input any details about the physical
conditions in the process zone. The technique can be implemented also in three dimensions, although we feel 
that this should wait for some increase in computational power before this can be implemented on a desktop
computer.

We want to indicate a few important direction along which the model can be applied, and in which we have started
some work. First of all, all simulations in the present paper were done in the overdamped regime, where
dynamical effects are absent. This may be a reasonable choice for the cases in which the cracks are known to grow
quasistatically. This may include crack propagation under slowly varying external conditions, as for instance non-uniform
thermal stresses. For cases in which dynamical effect are expected to be important the full dynamical equations
have to be implemented. Preliminary work shows that indeed the implementation of the inertial dynamics gives rise
(under appropriate conditions) to well known phenomena such as crack oscillation and bifurcation.
We expect to report about this soon.
Another possible interesting application of the model concerns the determination of minimum energy configuration
of cracks. In fact, as a result of regularization, our cracks are in principle able to shift laterally, in
addition to extend from its tips. This effect is not observed in the simulations presented here as it occurs
typically in much longer time scales than the one we were interested in, but it can be enhanced under particular
conditions. This shifting of cracks is driven by the tendency of the system to minimize its energy, and then it
provides a tool to study cases in which the minimum energy configuration has a physical meaning.
An example of this kind of application has been presented in \cite{jaglaPRE69}

As a final consideration, we stress that we are describing fracture in a 
continuum model of a brittle material
as a non-linear elastic process.
Due to the simplicity of the model, the influence of microscopic details at the process zone have only 
very few places were to leak in the present formalism. One point where this may happen is in the form
of the interpolation function between linear elasticity and broken material regime.
Based on very recent findings,\cite{nature04} we expect that particular changes in the form of this 
function may give rise to different phenomenological 
behavior of crack propagation.

\section{Acknowledgments}

We thank S. R. Shenoy for useful comments and discussions.

\newpage

\begin{figure}
\centerline{\epsfxsize=8.cm \epsfbox{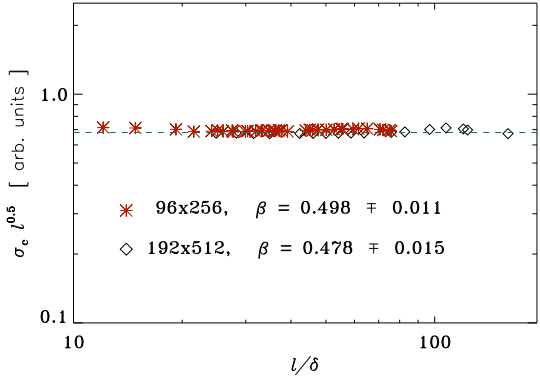}}
\caption{(Color online)
Critical remote stress for the propagation of straight  cracks of different lengths $l$ (with $\delta$
being the lattice discretization), placed along the 
$x$ direction, for two different 
system sizes (a value of $\mu=0.5B$ is used in all numerical simulations presented). 
Results are for the model without regularization ($\alpha=0$). The exponents
to fit the Griffith's law ($\sigma_c\sim l^{-\beta}$, with $\beta=1/2$) are indicated.
  }
\label{f3}
\end{figure}

\begin{figure}
\centerline{\epsfxsize=8.cm \epsfbox{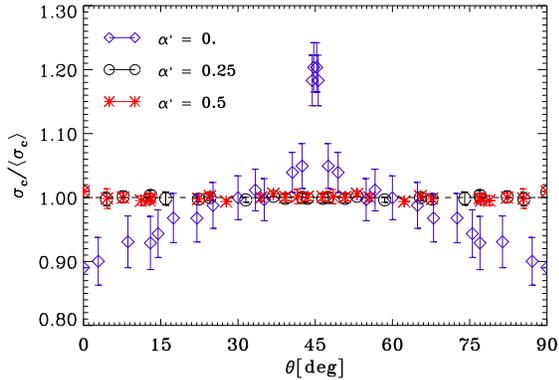}}
\caption{(Color online) Critical remote stress $\sigma_c$ for different values of $\alpha'\equiv\alpha/2B\delta^2$,
 normalized by the mean values, 
$\langle\sigma_c\rangle=0.2436B$ (for $\alpha'=0$), $\langle\sigma_c\rangle=0.9254B$ ($\alpha'=0.25$),
 $\langle\sigma_c\rangle=1.0776B$ ($\alpha'=0.5$),
for the propagation of cracks of length $l_0=40\delta$ 
placed at different angles with respect
to the numerical mesh ($256\times256$). The strong dependence with angle in the $\alpha=0$ becomes barely 
noticeable after the 
inclusion of regularization.}
\label{f4}
\end{figure}

\begin{figure}
\centerline{\epsfxsize=8.cm \epsfbox{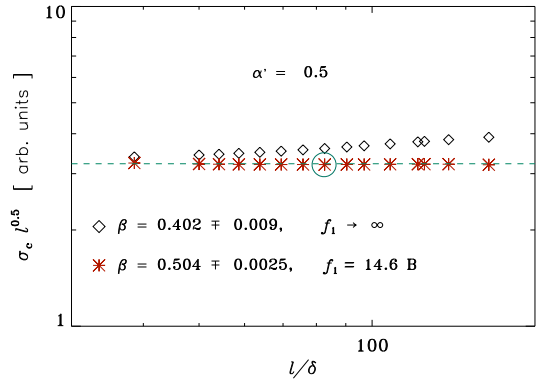}}
\caption{(Color online) Critical remote stress for the propagation of  straight cracks of different lengths $l$, obtained
including into the model the regularization terms, $\alpha'=0.5$. An exponent lower than $0.5$ is
obtained with infinite $f_1$ but a $\beta=0.5$ exponent is well fitted with a finite $f_1$ value. System
size: $192\times512$. }
\label{f7}
\end{figure}

\begin{figure}
\centerline{\epsfxsize=8.cm \epsfbox{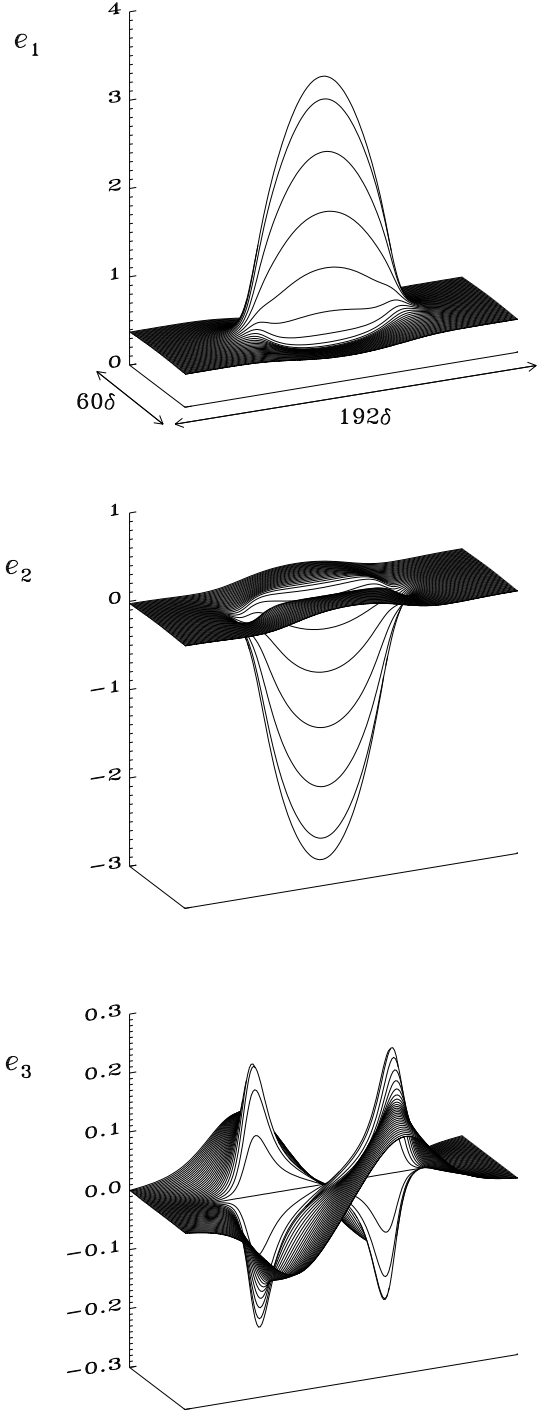}}
\caption{The full profiles for the three variable $e_1$, $e_2$, and $e_3$ of a crack with
$l=83\delta$, at the
critical stress $\sigma_c=0.7 B$, corresponding to the encircled point in Fig. \ref{f7}.}
\label{f8}
\end{figure}

\begin{thebibliography}{21}

\bibitem{reviews}H. Emmerich, {\it The diffuse Interface Approach in Condensed Matter}, Springer, Berlin (2003);
L.-Q. Chen, Annu. Rev. Mater. Sci. {\bf 32}, 113 (2002).

\bibitem{aronson}I. S. Aranson, V. A. Kalatsky, and V. M. Vinokur, Phys. Rev. Lett. {\bf 85} 118 (2000).

\bibitem{karma}A. Karma, D. A. Kessler, and H. Levine, Phys. Rev. Lett. {\bf 87}, 045501 (2001).

\bibitem{sethna}L. O. Eastgate, J. P. Sethna, M. Rauscher, T. Cretegny, 
C.-S. Chen, and C. R. Myers, Phys. Rev. E {\bf 65}, 036117 (2002).

\bibitem{kacha} The theory presented by Y. M. Jin, Y. U. Wang, and A. G. Khachaturyan [Philos. Mag. {\bf 83},
1587 (2003)] uses a phase field of tensorial nature, and then is close to the one we present here. However, it
seems to be applicable only when there is a finite number of cleavage planes in the material.

\bibitem{karma04} A. Karma and A. E. Lobkovsky, Phys. Rev. Lett. {\bf 92}, 245510 (2004), 
H. Henry and H. Levine,   Phys. Rev. Lett. {\bf 93}, 105504 (2004).

\bibitem{kessner} K. Kassner, Ch. Misbah, J. M\"{u}ller, J. Kappey, and P. Kohlert,
Phys. Rev. E {\bf 63}, 036117 (2001).

\bibitem{sharpin}E. A. Brener and R. Spatschek,   Phys. Rev. E {\bf 67}, 016112 (2003).
 
\bibitem{nota0}That is why we prefer to call our approach a diffuse interface one. We reserve the term
`phase field model' for cases where there are additional fields.

\bibitem{nota1}We do not include at present the non-linear part of the strain tensor 
($\sim\frac{\partial u_k}{\partial x_i}\frac{\partial u_k}{\partial x_j}$), and then in its present form
our formalism does not properly describe problems in which finite relative rotations of different parts
of the body occur.

\bibitem{shenoy}T. Lookman, S. R. Shenoy, K. O. Rasmussen, A. Saxena, and A. R. Bishop, Phys. Rev. B {\bf 67},
024114 (2003); 
S. R. Shenoy, T. Lookman, A. Saxena, and A. R. Bishop, Phys. Rev. B {\bf 60}, R12537
(1999).

\bibitem{kartha}S. Kartha, J. A. Krumhansl, J. P. Sethna, and L. K. Wickham, Phys. Rev. B {\bf 52}, 803
(1995).

\bibitem{stv}D. S. Chandrasekharaiah and L. Debnath, {\em Continuum Mechanics}, (Academic, San Diego, 1996).

\bibitem{nota}We prefer to work in terms of the three variables $e_1$, $e_2$, and $e_3$, and enforce the compatibility
condition by using a Lagrange multiplier. Notwithstanding, it should be mentioned that another possibility is 
to use only two truly independent variables, 
expressing explicitly  the third one in terms of the other two using the compatibility equation (\ref{stv}). 
In that case we would obtain a non-local, long range interaction between the two independent variables, as explained in 
Refs. \cite{shenoy} and \cite{kartha}.


\bibitem{nature}We expect that the exact form of the expression we use to interpolate between the quadratic
dependence at low $\varepsilon$ and the constant value at large $\varepsilon$ captures in the macroscopic model
some relevant details of the material at the microscopic scale (see {\protect {\cite{nature04}}}).

\bibitem{broberg}K. B. Broberg, {\it Cracks and Fracture}, (Academic, San Diego, 1999).

\bibitem{marchenko}E. A. Brener and V. I Marchenko, Phys. Rev. Lett {\bf 81}, 5141 (1998); 
R. Spatschek and E. A. Brener, Phys. Rev. E {\bf 64}, 046120 (2001).

\bibitem{fk}P. M. Chaikin and T. C. Lubensky, {\em Principles of Condensed Matter
Physics}, (Cambridge University Press, 1995).

\bibitem{prob}E. E. Gdoutos, {\em Problems of mixed mode crack propagation},
 (Martinus Nijhoff Publishers, 1984).

\bibitem{nature04}M. J. Buehler, F. F. Abraham, and H. Gao, Nature {\bf 426}, 141 (2003).

\bibitem{jaglaPRE69}E. A. Jagla, Phys. Rev. E {\bf 69}, 056212 (2004).

\end{thebibliography}
\end{document}